\documentclass[11pt,twoside]{article}

\usepackage{asp2006}
\usepackage{epsf}
\usepackage{psfig}
\usepackage{lscape}
\usepackage{graphicx}

\begin{document}
\title{Coronal Loop Models and Those Annoying Observations!}
\author{James A. Klimchuk}
\affil{NASA Goddard Space Flight Center, Code 671, Greenbelt, MD
20771, USA}

\begin{abstract}
It was once thought that all coronal loops are in static
equilibrium, but observational and modeling developments over the
past decade have shown that this is clearly not the case.  It is now
established that warm ($\sim 1$ MK) loops observed in the EUV are
explainable as bundles of unresolved strands that are heated
impulsively by storms of nanoflares.  A raging debate concerning the
multi-thermal versus isothermal nature of the loops can be
reconciled in terms of the duration of the storm.  We show that
short and long storms produce narrow and broad thermal
distributions, respectively. We also examine the possibility that
warm loops can be explained with thermal nonequilibrium, a process
by which steady heating produces dynamic behavior whenever the
heating is highly concentrated near the loop footpoints. We conclude
that this is not a viable explanation for monolithic loops under the
conditions we have considered, but that it may have application to
multi-stranded loops.  Serious questions remain, however.
\end{abstract}

\section{Introduction}

The unusual title of this paper is meant to indicate the emotional
aspects of being a coronal loops modeler.  Whenever we start to feel
confident that the problem is solved, new observations come along
and force us to modify our thinking.  It can be frustrating, but it
is also very rewarding when we gain improved physical understanding
of this fascinating phenomenon. The coronal loops problem is an
outstanding example of how the greatest progress is made when
observation and theory work together, one feeding off of the other.

The loops problem can be thought of as a puzzle, with the pieces of
the puzzle being observational constraints.  The goal is to fit the
pieces together into a physically consistent picture (there may be
more than one solution).  Five key pieces are:  (1) density, (2)
lifetime, (3) thermal distribution, (4) flows, and (5) intensity
profile. For the density, we are particularly interested in how the
observed density compares with the density that is expected for
static equilibrium. Thermal distribution refers to whether and how
the temperature varies over the loop cross section, i.e., across the
loop axis, and intensity profile refers to the variation of
brightness along the loop axis.

For many years, our picture of coronal loops was relatively simple
and the puzzle seemed easy to solve.  The observational constraints
came primarily from soft X-ray (SXR) observations of hot ($> 2$ MK)
loops. These loops were found to be long-lived
\citep[e.g.,][]{pk95jk} and to satisfy static equilibrium scaling
laws \citep[e.g.,][]{rtv78jk,kt96jk}.  The most straightforward
explanation was that these loops are heated in a steady fashion.

The picture became much more confused with new observations of warm
($\sim 1$ MK) loops made in the EUV by {\it SOHO}/EIT and {\it
TRACE}. These warm loops can appear to occupy the same volume as hot
loops---though not necessarily at the same time---but their
properties are fundamentally different. Besides the obvious
temperature difference, EUV loops are over dense relative to static
equilibrium, they have super-hydrostatic scale heights, and they
have exceptionally flat temperature profiles when measured with the
filter ratio technique
\citep[e.g.,][]{aetal99jk,letal99jk,asa01jk,wwm03jk}. These loops
are clearly not in static equilibrium.

This paper describes the logical progression that has been followed
by the loops community in attempting to explain the observations,
especially those of the more challenging EUV loops.  We represent
this progression with the flowchart in Figure 1, which is in many
ways a recent history of how the discipline has evolved.

\begin{figure}[!ht]
\centering\includegraphics[width=10.0cm, angle=-90, trim = 50 130 20
0]{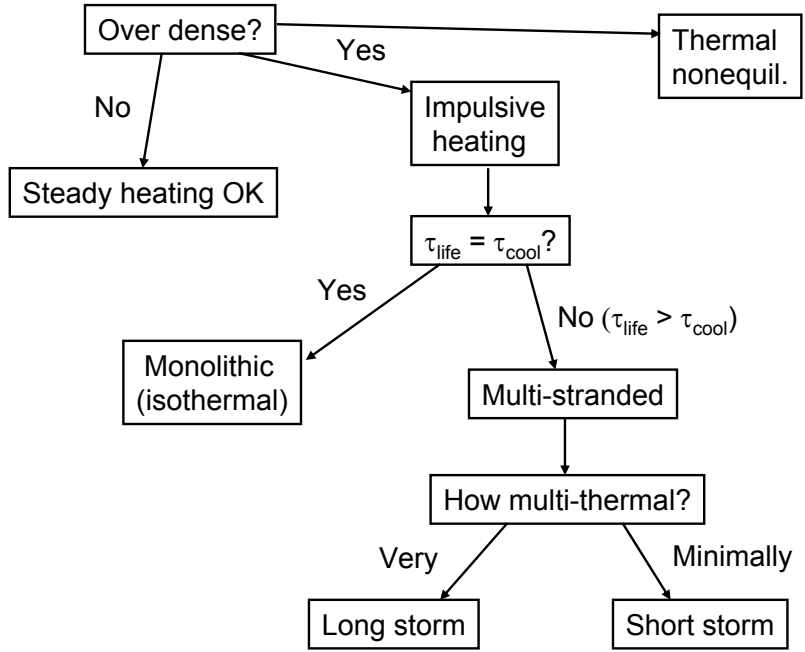}
%\plotone{klimchuk_fig1.eps}
%\psfig{figure=klimchuk_fig1.ps,bbllx=0pt,bblly=0pt,bburx=720pt,bbury=540pt,width=11cm}
%\centering\includegraphics[width=11.0cm]{klimchuk_fig1.ps}
%\includegraphics[width=11.0cm, angle=-90, trim = 50 130 20 0]{klimchuk_fig1.pdf}
\caption{Flow chart showing the logical progression used to infer
the physical nature and heating of a coronal loop.  Some boxes
indicate observational questions and others indicate conclusions
that are drawn from the answers.}
\end{figure}

\section{Density}

Suppose we wish to investigate an observed loop.  We can start by
asking the question ``Is the loop over dense relative to static
equilibrium?"  Given the observed temperature and length, static
equilibrium theory predicts a unique density.  We want to know
whether the observed density is larger than this value? If it is
not, and if the loop does not evolve rapidly, then steady heating is
a possible, though not unique, explanation. This was essentially
where things stood through the {\it Yohkoh} mission in the 1990's.

As we have already indicated, however, most EUV loops are indeed
over dense. This is indicated in Figure 2 \citep[reproduced from
][]{k06jk}, which reveals the physics of what is going on. The
figure shows the ratio of the radiative to conductive cooling times
plotted against temperature for a large sample of loops.  The warm
loops were observed by {\it TRACE}, and the hot loops were observed
by {\it Yohkoh}/SXT. The ratio of the cooling times is determined
from the measured temperature, density, and length according to
$\tau_{rad}/\tau_{cond} = T^4/(n^2L^2)$, although the power of
temperature depends weakly on the radiative loss function and is
slightly different in different temperature regimes.

\begin{figure}[!ht]
\centering\includegraphics[width=11.0cm]{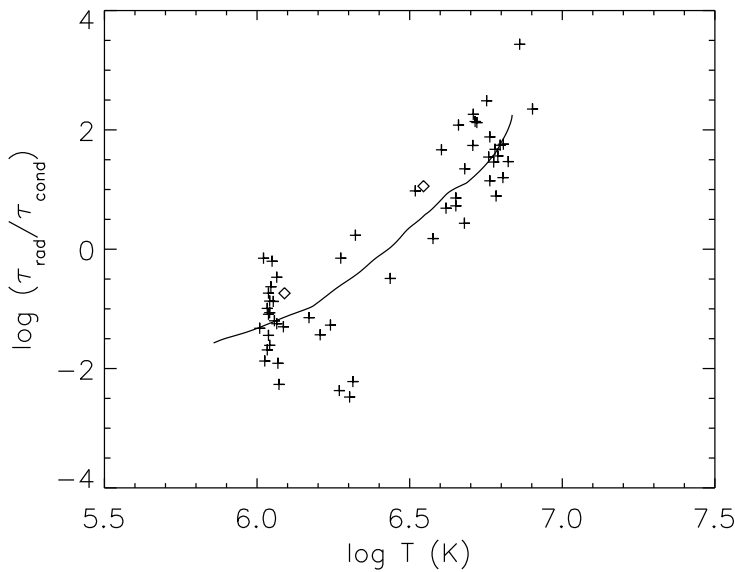}
\caption{Ratio of radiative to conductive cooling times versus
temperature for many observed loops.  Solid line is the cooling
track of an impulsively heated loop strand simulation.  From
Klimchuk (2006).}
\end{figure}

Coronal energy losses from radiation and thermal conduction are
comparable for loops that are in static equilibrium \citep{vau79jk},
and such loops would fall along a horizontal line near 0 in the
plot.  Loops that lie above the line are under dense, and loops that
lie below the line are over dense. The observed loops follow a clear
trend ranging from hot and under dense in the upper-right to warm
and over dense in the lower-left. Note, however, that the densities
used for the cooling time ratios were measured using emission
measures and loop diameters and assuming a filling factor of unity,
$n = [EM / (df)]^{1/2}$, so they are lower limits. Smaller filling
factors would shift the points downward in the plot. Thus, the hot
loops could be in static equilibrium, and the warm loops could be
even more over dense than indicated.

It is abundantly clear that static equilibrium cannot explain warm
loops.  An explanation relying on steady end-to-end flows is also
not viable \citep{pkm04jk}.  Thermal nonequilibrium is a possibility
that we return to later.  The most promising explanation for the
observed over densities of warm loops is implusive heating.  This
can also explain the under densities of hot loops, if they are real.
The solid curve that fits the points so well in Figure 2 is the
evolutionary track from a 1D hydrodynamic simulation of a loop that
has been heated impulsively by a nanoflare. Cooling begins at the
upper-right end of the track and progresses downward and to the
left. The early stages are dominated by thermal conduction and are
characterized by under densities, while the late stages are
dominated by radiation and are characterized by over densities. The
ability of nanoflare models to reproduce the observed densities of
loops is well established
\citep{k02jk,wwh02jk,wwm03jk,ck04jk,k06jk}.

\section{Lifetime}

If a loop is heated impulsively, then we might expect it to exist
for approximately a cooling time (combining the effects of
conduction and radiation), as determined from the observed
temperature, density, and length.  This is the next question in the
flowchart. If the lifetime and cooling time are similar, we can
conclude that the loop is a monolithic structure that heats and
cools as a homogeneous unit, with uniform temperature over the cross
section. Observations show that this is not case, however. The vast
majority of loops live longer than a cooling time and sometimes much
longer \citep[e.g.,][]{wws03jk,lkm07jk}. If these loops contain
cooling plasma, then they cannot be monolithic. Rather, they must be
bundles of thin, unresolved strands that are heated impulsively at
different times.  Although each strand cools rapidly, the composite
bundle appears to evolve slowly \citep[e.g.,][]{wwm03jk}.
Multi-stranded bundles of this type can explain a number of observed
properties of warm loops:  over density, long lifetime,
super-hydrostatic scale height, and flat temperature profile.  They
can also explain the observed under density of hot loops.  Realizing
this was a time of rejoicing in the modeling community! But....

\section{Thermal Distribution}

An important prediction of the multi-strand model is that loops
should have multi-thermal cross-sections.  Since the unresolved
strands are heated at different times, they will be in different
stages of cooling and out of phase with each other.  A critical
question became ``Are loops multi-thermal?"  An intense debate
ensued and continues to this day.  Some have answered with a
resounding yes \citep[the ``Schmelz camp," e.g.,][]{sm06jk} and
others have answered with a resounding no \citep[the ``Aschwanden
camp," e.g.,][]{an05jk}. As we now demonstrate, however, it is not
especially useful to phrase the multi-thermal question in a way that
requires a binary response.

Imagine that a loop bundle is heated by a ``storm" of nanoflares
that occur randomly over a finite window in time.  It is easy to see
that the range of strand temperatures that are present at any given
moment depends on the duration of the storm. For a very short storm,
all of strands will be heated at about the same time and will cool
together.  The instantaneous thermal distribution of the loop will
be narrow.  In contrast, a storm that lasts longer than a cooling
time will produce a much wider thermal distribution. Some strands
will have just been heated and will be very hot; others will have
cooled to intermediate temperatures; and still others will have had
time to cool to much lower temperatures. The flowchart in Figure 1
therefore asks the more meaningful question ``How multi-thermal is
the loop?" A broad thermal distribution implies a long-duration
nanoflare storm, and a narrow distribution implies a short-duration
storm. It now appears that the multi-thermal and isothermal camps
may both be correct.

The duration of the nanoflare storm also determines the lifetime of
the loop bundle, so the thermal width and lifetime will be closely
related. Figure 3 shows results for simulated nanoflare storms
lasting 500, 2500, and 5000 s, top to bottom. The left column has
light curves (intensity versus time) as would be observed in the 195
channel of {\it TRACE}, with sensitivity peaking near 1 MK.  The
right column has emission measure distributions, EM($T$) = $T \times
\!$DEM($T$) cm$^{-5}$, at the time of peak 195 intensity. Only the
coronal part of the loop is included; the transition region
footpoints are neglected. All three of the storms are comprised of
identical nanoflares that have triangular heating profiles lasting
500 s. They were simulated with our ``0D" hydro code EBTEL and are
the same as example 4 in \citet{kpc08jk}. In actuality, Figure 3 was
produced with only one simulation. The light curves and EM
distributions were constructed using sliding time windows that
correspond to the storm durations.

\begin{figure}[!ht]
\centering\includegraphics[width=13.cm, trim = 25 0 0
0]{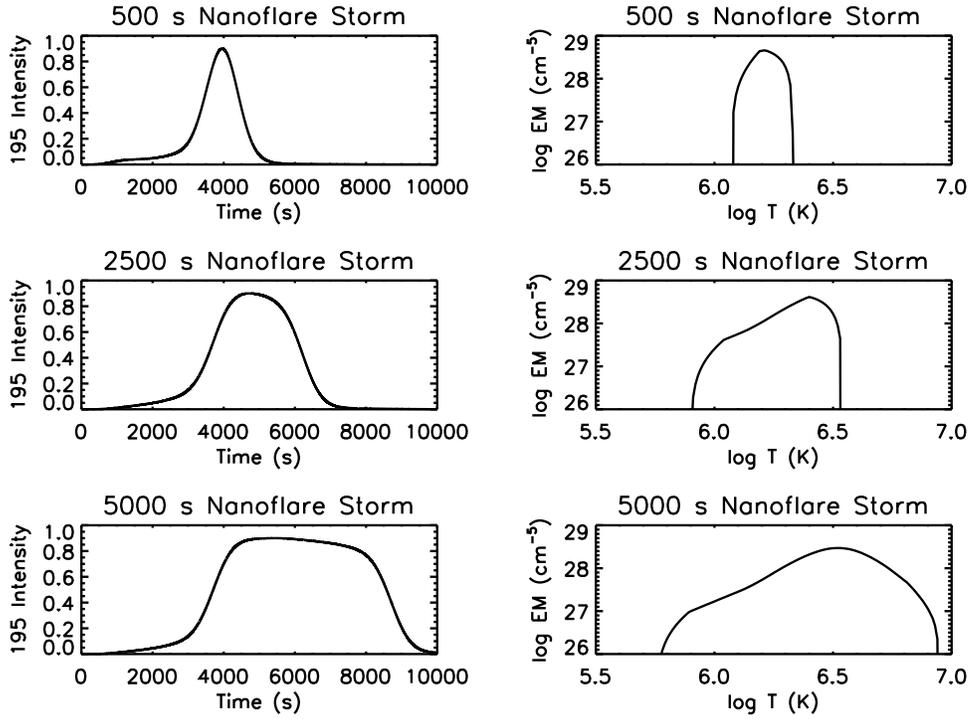}
\caption{Simulated 195 light curves (left) and
emission measure distributions (right) for nanoflare storms lasting
500, 2500, and 5000 s, top to bottom.  The instantaneous EM
distributions are from the times of peak 195 intensity (t = 3958,
4705, 5445 s for the three storms).}
\end{figure}

As expected, both the lifetime and thermal width increase as the
storms get longer.  The full widths at half maximum (FWHM) of the
light curves are 1098, 2579, and 5008 sec for the 500, 2500, and
5000 s storms, respectively.  The FWHM of the EM distributions are
0.13, 0.23, and 0.36 in $\log T$.  The full widths at the 1\% levels
are 0.24, 0.62, and 1.14 in $\log T$.

It may seem surprising at first that the EM distributions do not all
reach the same maximum temperature, since the nanoflares are the
same in all three storms.  This is {\it not} because individual
strands are reheated multiple times in the longer storms; all
strands are heated only once.  Rather, it is because the
distributions are from the time of peak 195 intensity.  In the short
duration storm, all of the strands have cooled appreciably by the
time the peak intensity is reached.  Had we chosen to plot the
distribution at an earlier time, it would still been narrow, but it
would be shifted to higher temperature.

\citet{wetal08jk} have made Gaussian fits to EM distributions
observed by {\it Hinode}/EIS.  They find a typical central
temperature of 1.4 MK and a typical Gaussian half width of 0.3 MK.
This corresponds to a FWHM in $\log T$ of roughly 0.24, which by
Figure 3 implies a 195 lifetime of roughly 2500 s. Although Warren
et al. did not measure the lifetimes of their loops, this value is
consistent with the small number of 195 lifetimes that have been
reported for other cases \citep{ww05jk,uww06jk}.  To our knowledge,
there does not exist a single published example where both the
thermal width and lifetime have been measured for the same loop.
Making such measurements should be a high priority.  It is a crucial
consistency check of the nanoflare concept. Density measurements
should be made at the same time.

\section{Very Hot and Very Faint Plasma}

The nanoflare model makes two observational predictions in addition
to the ones we have already discussed.  First, it predicts that
small amounts of very hot ($> 5$ MK) plasma should be present.
Figure 4 shows two examples of long (infinite) duration storms, one
comprised of relatively weak nanoflares and the other comprised of
nanoflares that are ten times stronger. The solid curve in each case
is the EM distribution for the whole loop, while the dashed and
dot-dashed curves are the contributions from the coronal section and
footpoints, respectively. We see that the EM of the hottest plasma
is 1.5-2 orders of magnitude smaller than that of the most prevalent
plasma. The reason is two-fold. First, the initial cooling after the
nanoflare has occurred is very rapid, so the hottest plasma persists
for a relatively brief period. Second, the densities are low during
this early phase, because chromospheric evaporation has only just
begun to fill the loop strand with plasma.

\begin{figure}[!ht]
\psfig{figure=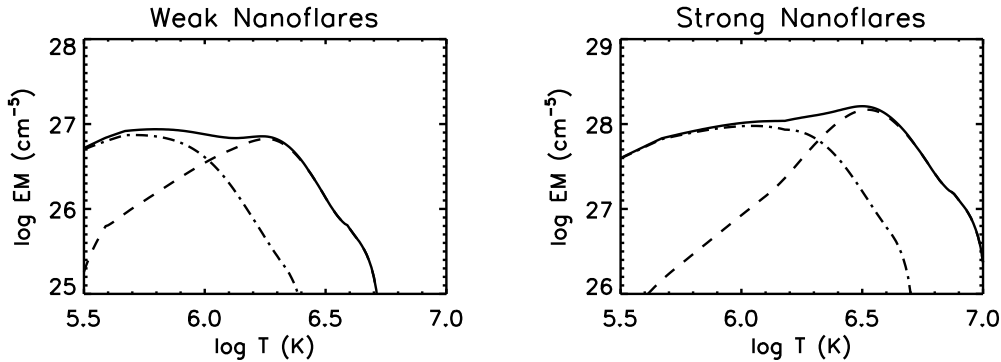,bbllx=80pt,bblly=540pt,bburx=470pt,bbury=720pt,width=11cm}
%\centering\includegraphics[width=11.cm, trim = 80 540 0 60]{klimchuk_fig4.ps}
%\includegraphics[width=15.cm]{klimchuk_fig4.pdf}
\caption{Emission measure distributions for long (infinite) duration
nanoflare storms comprised of weak (left) and strong (right)
nanoflares: coronal section (dashed), transition region footpoints
(dot-dashed), and whole loop (solid).}
\end{figure}

As a consequence of the small emission measures, the intensities of
hot spectral lines and channels are predicted to be very faint.  The
intensities may be reduced still further by ionization
nonequilibrium effects \citep{bc06jk,ro08jk}. Low levels of
super-hot emission have nonetheless been detected recently by the
{\it CORONAS}, {\it RHESSI}, and {\it Hinode} missions
\citep{zetal06jk,m09jk,pk09jk,ketal09jk}.  In particular, EM
distributions inferred from multi-filter XRT observations of two
active regions suggest that the distributions may have two distinct
components \citep{setal09jk,retal09jk}. The implications are
considerable, since this would rule out a simple power-law energy
distribution for the responsible nanoflares. Detailed modeling is
now underway.

\section{Flows}

High-speed upflows that reach or exceed 100 km s$^{-1}$ are
predicted during the early evaporation phase of a nanoflare event.
Depending on the geometry of the observations, these can produce
highly blue-shifted emission.  The emission will be very faint,
however, for the reasons given above.  A composite spectral line
profile from a bundle of unresolved strands will be dominated by the
weakly red-shifted emission produced during the much longer
radiative cooling phase, when the plasma slowly drains and condenses
back onto the chromosphere.  Signatures of evaporation take the form
of blue wing enhancements on this main component \citep{pk06jk}.
They can be very subtle, and they only appear in lines that are well
tuned to the temperature of the evaporating plasma. Significantly
hotter and cooler lines are not expected to show evidence of
evaporation.

We have performed sit-and-stare observations with {\it Hinode}/EIS
and find blue wing asymmetries in Fe XVII ($T \approx 5$ MK) similar
to those predicted by our nanoflare models \citep{pk06jk}. The
measurements are very challenging, however, due to the faint nature
of the line. \citet{hetal08jk} also report blue-wing asymmetries
that are suggestive of nanoflares.

\section{Thermal Nonequilibrium}

We have worked our way down the flowchart of Figure 1 and concluded
that the observed properties of many loops can be explained by
storms of nanoflares occurring within bundles of unresolved strands.
There remains the possibility, indicated in the upper right, that
many loops can also be explained by thermal nonequilibrium.  We
consider this possibility now.

Thermal nonequilibrium is a fascinating phenomenon in which dynamic
behavior is produced by perfectly steady heating
\citep{ak91jk,ketal01jk,mph04jk,kak06jk}. No equilibrium exists if
the steady heating is sufficiently highly concentrated near the loop
footpoints. Instead, the loop goes through periodic convulsions as
it searches for a nonexistent equilibrium. Cold, dense condensations
form, slide down the loop leg, and later reform in a cycle that
repeats with periods of several tens of minutes to several hours.

We have recently explored whether thermal nonequilibrium can explain
the observed properties of EUV loops \citep{kk09jk}.  We first
considered a monolithic loop, which we simulated with our 1D hydro
code ARGOS \citep{skaetal99jk}.  The code uses adaptive mesh
refinement, which is critical for resolving the thin transition
regions that exist on either side of the dynamic condensations.  We
imposed a steady heating that decreases exponentially with distance
from both footpoints.  The heating scale length of 5 Mm is
one-fifteenth of loop halflength. We introduced a small asymmetry by
making the amplitude of the heating on the right side only 75\% that
on the left.

Figure 5 shows the evolution of temperature, density, and intensity
as would be observed in the 171 channel of {\it TRACE}.  These are
averages over the upper 80\% of the loop.  The behavior is typical
of the several cycles that we simulated.  The loop is visible in 171
for only about 1000 s.  This is a factor of 2-4 shorter than
observed lifetimes \citep{ww05jk,uww06jk}.  A more serious problem
is the distribution of emission along the loop (the intensity
profile), which disagrees dramatically with observations. Figure 6
shows 171 intensity and temperature as a function of position along
the loop at $t = 5000$ s.  The emission is strongly concentrated in
transition region layers at the loop footpoints ($s = 45$ and $203$
Mm) and to either side of a cold condensation at $s = 163$ Mm. In
stark contrast, most observed 171 loops have a fairly uniform
brightness along their length.

\begin{figure}[!ht]
\centering\includegraphics[width=10.cm]{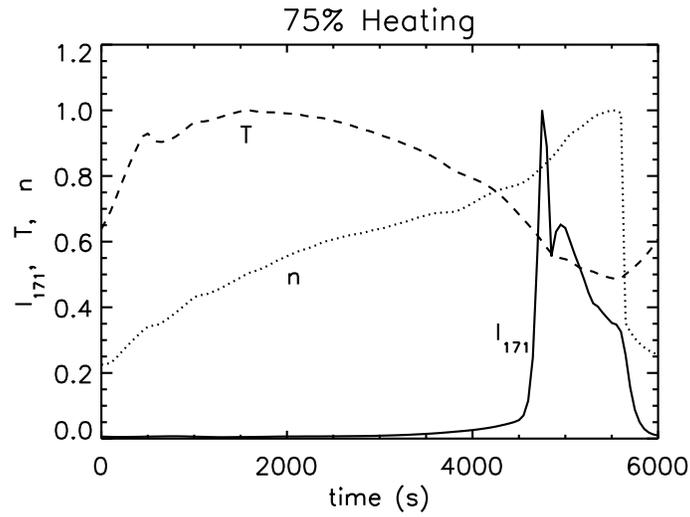}
\caption{Evolution of temperature (dashed), density (dotted), and
171 intensity (solid) for a monolithic loop undergoing thermal
nonequilibrium.  All quantities are normalized.  The steady heating
is 75\% as strong in the right leg as in the left.}
\end{figure}

\begin{figure}[!ht]
\centering\includegraphics[width=10.cm]{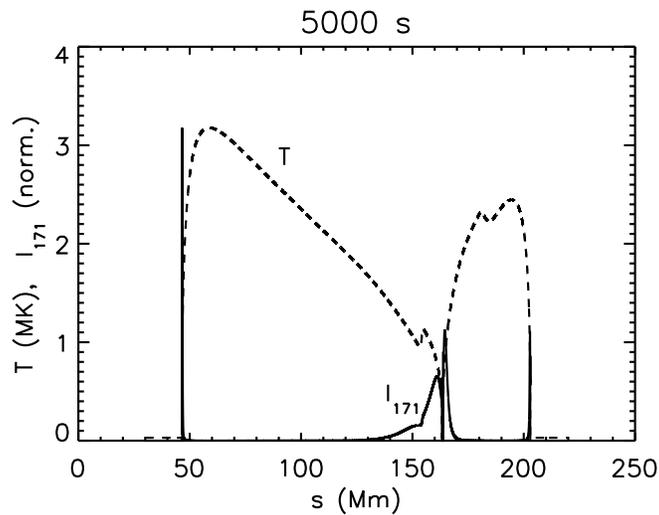}
\caption{Temperature (dashed, MK) and 171 intensity (solid,
arbitrary units) as a function of position along the loop at $t =
5000$ s in the simulation of Figure 5.}
\end{figure}

The maximum temperature in the loop is 4.4 MK and occurs before the
condensation forms. We performed another simulation with a reduced
heating rate that has a maximum temperature of only 1.8 MK. Neither
the light curve nor the intensity profile are consistent with
observations.  We conclude that EUV loops are not monolithic
structures undergoing thermal nonequilibrium, at least not under
conditions that lead to cold condensations.  We note, however, that
\citet{metal08jk} report a different type of nonequilibrium
behavior. One prominent loop  in their 3D simulation of an active
region exhibits a cooling and heating cycle, but the temperature
never drops to the point where a condensation forms. The reasons for
the differing behavior are yet to be understood.  Whether the loop
has properties matching observed loops (density, lifetime, thermal
width) is unknown.

The simulation of Figures 5 and 6 may nonetheless have some
relevance to the Sun. The condensation falls onto the right
footpoint at $t = 5600$ s. Falling condensations have been seen in
the C IV channel of {\it TRACE} \citep{s01jk}.  They are relatively
rare, however, and occur in only a small fraction of loops.

We next considered the possibility of a multi-stranded loop bundle
in which the individual strands undergo thermal nonequilibrium in an
out-of-phase fashion. To approximate such a loop, we performed two
additional simulations, similar to the first but with heating
imbalances of 50\% and 90\% instead of 75\%. We then averaged all
three simulations in time and added them together along with their
mirror images to form a composite loop.  The resulting 171 intensity
profile is shown in Figure 7. It is reasonably uniform except for
the very intense spikes at the footpoints (note the logarithmic
scale). A more realistic loop bundle with a wider variety of heating
imbalances would be even more uniform.  We tentatively conclude that
the intensity profile is consistent with observations, although we
are concerned because bright 171 moss emission is generally observed
at the footpoints of SXR loops rather than the footpoints of EUV
loops.

\begin{figure}[!ht]
\centering\includegraphics[width=10.cm]{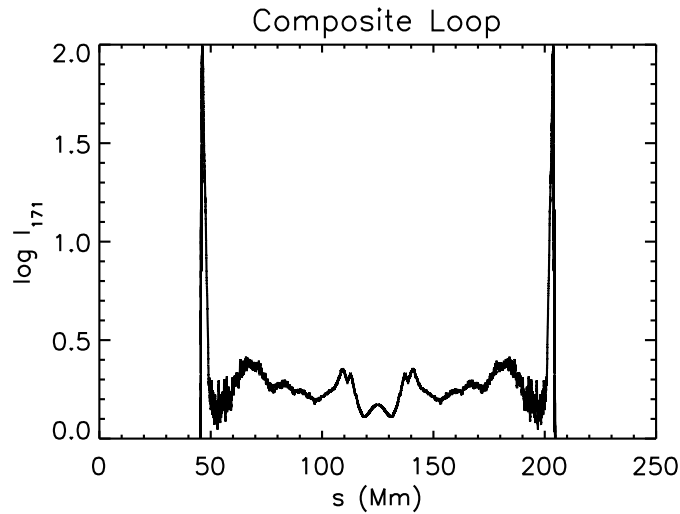}
\caption{Logarithm of 171 intensity as a function of position along
a composite loop bundle comprised of individual strands undergoing
thermal nonequilibrium. See text for details.}
\end{figure}

\begin{figure}[!ht]
\centering\includegraphics[width=10.cm]{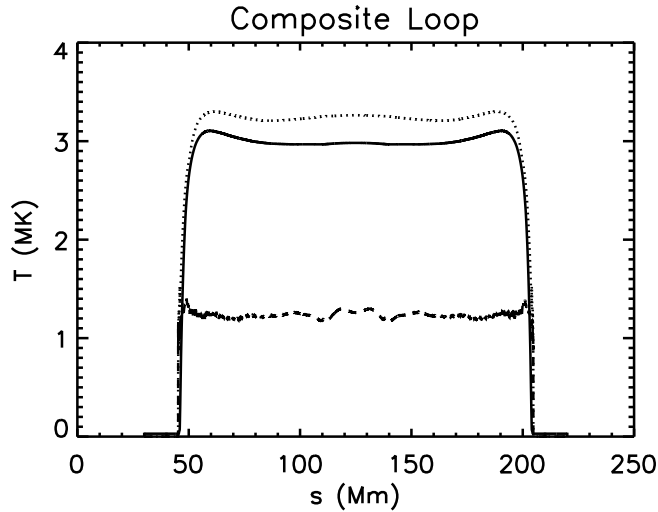}
\caption{Temperature as a function of position along the composite
loop bundle of Figure 7.  Solid is the actual mean temperature,
while dashed and dotted are the temperatures inferred from {\it
TRACE} and {\it Yohkoh}/SXT filter ratios, respectively.}
\end{figure}

Figure 8 shows three temperature profiles for the composite loop:
the average of the actual temperatures in the individual strands
(solid), the temperature that would be inferred from {\it TRACE}
171/195 intensity ratios (dashed), and the temperature that would be
inferred from {\it Yohkoh}/SXT Al12/AlMg intensity ratios (dotted).
They are different because {\it Yohkoh}/SXT is more sensitive to the
hotter plasma and {\it TRACE} is more sensitive to the warmer
plasma. Notice that the profiles is very flat. This is a well-know
property of EUV loops. We have also inferred densities from the
simulated {\it TRACE} observations using exactly the same procedure
that was used for the real loops in Figure 2. The model loop is over
dense by a factor of 23, consistent with observed values.  We have
repeated this excise using reduced heating in the strands and find
that the over density is a factor of 10 in this case.

Although there is some reason for encouragement, it is not obvious
that bundles of unresolved strands undergoing thermal nonequilibrium
can explain all the salient properties of observed EUV loops.
Reproducing the lifetimes is especially challenging. The
condensations in the different strands must be sufficiently out of
phase to give a uniform intensity profile, but they cannot be so out
of phase as to produce a composite loop lifetime longer than 1 hour.
Even if the phasing is correct for one condensation cycle, it is
likely to be incorrect for subsequent cycles because the interval
between condensations depends on both the amplitude of the heating
and its left-right imbalance.  The imbalance determines the location
where the condensation forms, and it must be appreciably different
among the strands in order to get a uniform intensity profile. Note
that the results shown in Figures 7 and 8 make use of temporal
averages over complete cycles, and therefore the lifetime of the
equivalent loop bundle is effectively infinite.

\section{Conclusions}

We have described how a combination of observational and modeling
work has led to the conclusion that warm ($\sim 1$ MK) EUV loops can
be explained as bundles of unresolved strands that are heated by
storms of nanoflares.  Static equilibrium is out of the question.
The observed lifetimes and thermal distributions of the plasma
indicate that the storms last for typically 2-4$\times10^3$ s.
Additional support for this picture is provided by the shapes of hot
spectral line profiles and by the observation that line intensities
peak at slightly later times for lines of progressively cooler
temperature \citep{uwb09jk}.  Also, there is now good evidence for
very hot and very faint plasma, as predicted by the nanoflare
models.

It is not clear whether most hot ($> 2$ MK) SXR loops are also
heated by nanoflares.  If they are, the storms must be long duration
in order to explain the observed lifetimes. The loops would then be
expected to have co-spatial EUV counterparts, and it is not obvious
that they do.  One possibility is that the frequency of nanoflares
is much higher in long-lived SXR loops, so that the plasma in a
strand never cools to EUV temperatures before being reheated. It is
worth noting that virtually all of the proposed coronal heating
mechanisms predict impulsive energy release on individual magnetic
field lines \citep{k06jk}.

We considered the possibility that EUV loops can be explained by
thermal nonequilibrium. We concluded that this is not a viable
mechanism for monolithic loops under the conditions we have
considered---although the results of \citet{metal08jk} are very
intriguing---but that it may have application in multi-stranded
bundles. Serious questions remain that require further
investigation.

We close by pointing out that distinct loops are only one component
of the corona and that the diffuse component contributes at least as
much emission.  It is not generally appreciated that the intensity
of EUV and SXR loops is typically much less than that of the
background (of order 10-40\%). The diffuse component may also be
made up of individual strands, but we must explain why the strands
have a higher concentration in loops.

\acknowledgements I am very pleased to acknowledge useful
discussions with many people, but I especially wish to thank Spiros
Patsourakos, Harry Warren, and Judy Karpen, my collaborator on the
thermal nonequilibrium study that is being published here for the
first time. I benefited greatly from participation in the Coronal
Loops Workshop Series and the International Space Science Institute
team led by Susanna Parenti. Financial support came primarily from
the NASA Living With a Star program.

\end{document}